\documentclass[epj,referee]{svjour}
\usepackage{graphics}
\begin{document}
\title{The value of information in a multi-agent market model}
\subtitle{The luck of the uninformed}
\author{Bence T\'oth\inst{1,2}, Enrico Scalas\inst{1,3}, J\"urgen Huber\inst{4,5} \and Michael Kirchler\inst{5}
}                     
%
%
\institute{ISI Foundation - Torino, Italy, \and Department of Theoretical Physics, Budapest University of Technology and Economics - Budapest, Hungary \and Dipartimento di Scienze e Tecnologie Avanzate, East Piedmont University - Alessandria, Italy \and Yale School of Management - New Haven, USA\and Innsbruck University School of Management, Department of Banking and Finance - Innsbruck, Austria}
\date{Received: date / Revised version: date}
%
\abstract{
We present an experimental and simulated model of a multi-agent stock market driven by a double auction order matching mechanism.
Studying the effect of cumulative information on the performance of traders, we find a non monotonic relationship of net returns of traders as a function of information levels, both in the experiments and in the simulations. Particularly, averagely informed traders perform worse than the non informed and only traders with high levels of information (insiders) are able to beat the market. The simulations and the experiments reproduce many stylized facts of tick-by-tick stock-exchange data, such as fast decay of autocorrelation of returns, volatility clustering and fat-tailed distribution of returns. These results have an important message for everyday life. They can give a possible explanation why, on average, professional fund managers perform worse than the market index.
} 
\maketitle
\newpage
\newpage
\section{Introduction}
\label{intro}
"We live in an information society" is a commonly used phrase today. Education, knowledge and information are considered to be the most important ingredients to success in business. While we generally agree with this notion, we think that it does not always hold for financial markets. 70 years ago Cowles \cite{cowles.1933} was the first to find that the vast majority of stock market forecasters and fund managers are not able to beat the market. Subsequent studies by Jensen \cite{jensen.1968} and Malkiel \cite{malkiel.2003a,malkiel.2003b} confirmed this finding. On average about 70 percent of actively managed stock market funds are outperformed by the market, for bonds the number is even higher at 90 percent. Passive investment yields on average 1.5 percent per annum more than an actively managed fund \cite{malkiel.2003a}. How can we explain that the highly paid, professionally trained and, above all, well informed specialists managing these funds are not able to perform better than the market? 
The question whether more information is always good for market participants is highly relevant not only for fund managers, investment banks and regulators, but for every individual investor as well. 
In this paper we present results from experimental and simulation studies which allow improving our understanding of the relationship between information and investment success in markets. Our model features several innovations: First, our model is a multi-period model and therefore dynamic. It thereby overcomes one of the major weaknesses of earlier research relying only on static environments. Second, we use several information levels instead of only two used in most of the literature on the topic (e.g. Refs. \cite{grossman.stiglitz.1980,hellwig.1982,figlewski.1982,sunder.1992,scalas2005}). This is critical to go beyond the straightforward (and not surprising) result that insiders are able to outperform uninformed investors. As we will see the most interesting cases lie between these extremes. The averagely informed traders are the ones we are most interested in, as they exhibit underperformance in our experiments.

The paper is structured as follows: After the introduction, Section \ref{sec:1} presents the outline of our experiments, including the settings, information setup and results. Section \ref{Section3} presents the simulations with subchapters for the market mechanism, the information system, and trading strategies. Results from the simulations are presented in Section \ref{Section4}, and Section \ref{conclusions} concludes the paper.

\section{Outline of Experiments}
\label{sec:1}
The experiments we discuss here have been performed by two of us (J\"urgen Huber and Michael Kirchler) at the University of Innsbruck in 2004 with the participation of business students. To reduce statistical errors the experiments were repeated seven times with different subjects. 
For more details on the experimental setup and the results see Ref. \cite{kirchler.huber.2006}. 

\subsection{Settings of the Experiments}\label{exp_settings}
The experiments were based on a cumulative information system. Nine traders with different forecasting abilities were trading on a continuous double auction with limit orders and market orders. (Additional experiments were run with 20 traders distributed among five different information levels. The main result - a J-shaped distribution of returns - always emerged \cite{huber.2006}.) On the market a risky asset
(stock) and a risk free bond (cash) were traded. Any time, traders could enter a new limit order to the book or accept someone's limit order (realising a market order) with all trades fixed to unit volume. Each trader had a starting endowment of 1600 units in cash and 40 shares of stock (each worth 40 units  in the beginning). The experiment consisted of 30 periods each lasting 100 trading seconds. At the end of each period a risk free interest rate was paid on the cash held by the traders and dividends were paid based on the shares owned, with parameters set to let one period correspond to one month in real market. 
The dividend process (\(D(i)\)) was a random walk with Gaussian steps:
\begin{eqnarray}
D(i)=D(i-1)+0.0004 N(0,1)
\end{eqnarray}
with \(D(0)=0.2\), where \(N(0,1)\) is a normal distribution with zero mean and unit variance.
To achieve identical conditions, the same dividend process was used for all runs of the experiment.

\subsection{Information setup}\label{exp_info}
To value the shares, traders on the market got information about future dividends. The idea of Hellwig \cite{hellwig.1982} was extended to nine information levels: different levels of information correspond to different lengths of windows in which one can predict future dividends. Trader \textit{I1} knows the dividend for the end of the current period, trader \textit{I2} knows the dividends for the current and the next period, \dots, trader \textit{I9} for the current and the next eight periods \cite{huber.2006,huber.kirchler.sutter.2006}. This way we got a cumulative information structure of the market where better informed agents know future dividends earlier than less informed ones. Since the market trading consists of several periods (new information entering the market in each), the design implies that information trickles down through the market from the best informed to the broad public over time.\\
The information that traders obtain is the present value of the stock conditioned on the forecasting horizon (\(E(V|I_{j,k})\)). This is calculated using Gordon's formula, discounting the known dividends and assuming the last one as an infinite stream which is also discounted. \(E(V|I_{j,k})\) stands for the conditional present value of the asset in period \textit{k} for the trader with information level \textit{j} (\textit{Ij}).
\begin{equation}
E(V|I_{j,k})=\frac{D(k+j-1)}{r_{e}(1+r_e)^{j-2}}+\sum_{i=k}^{k+j-2}\frac{D(i)}{(1+r_{e})^{i-k}},
\label{eq_gordon}
\end{equation}

where \(r_{e}\) is the risk adjusted interest rate (\(E(\cdots|\cdots)\) denotes the conditional average).

Before the beginning of the experiment an information level from one to nine (\textit{I1,\dots,I9}) was randomly assigned to each trader which he/she kept for the whole session. There was one trader for each information level and this was public knowledge. At the beginning of each period new information was delivered to the traders depending on their level of information.

\subsection{Results of the Experiments}\label{exp_results}
\begin{figure}
\begin{center}
\resizebox{0.75\columnwidth}{!}{%
  \includegraphics{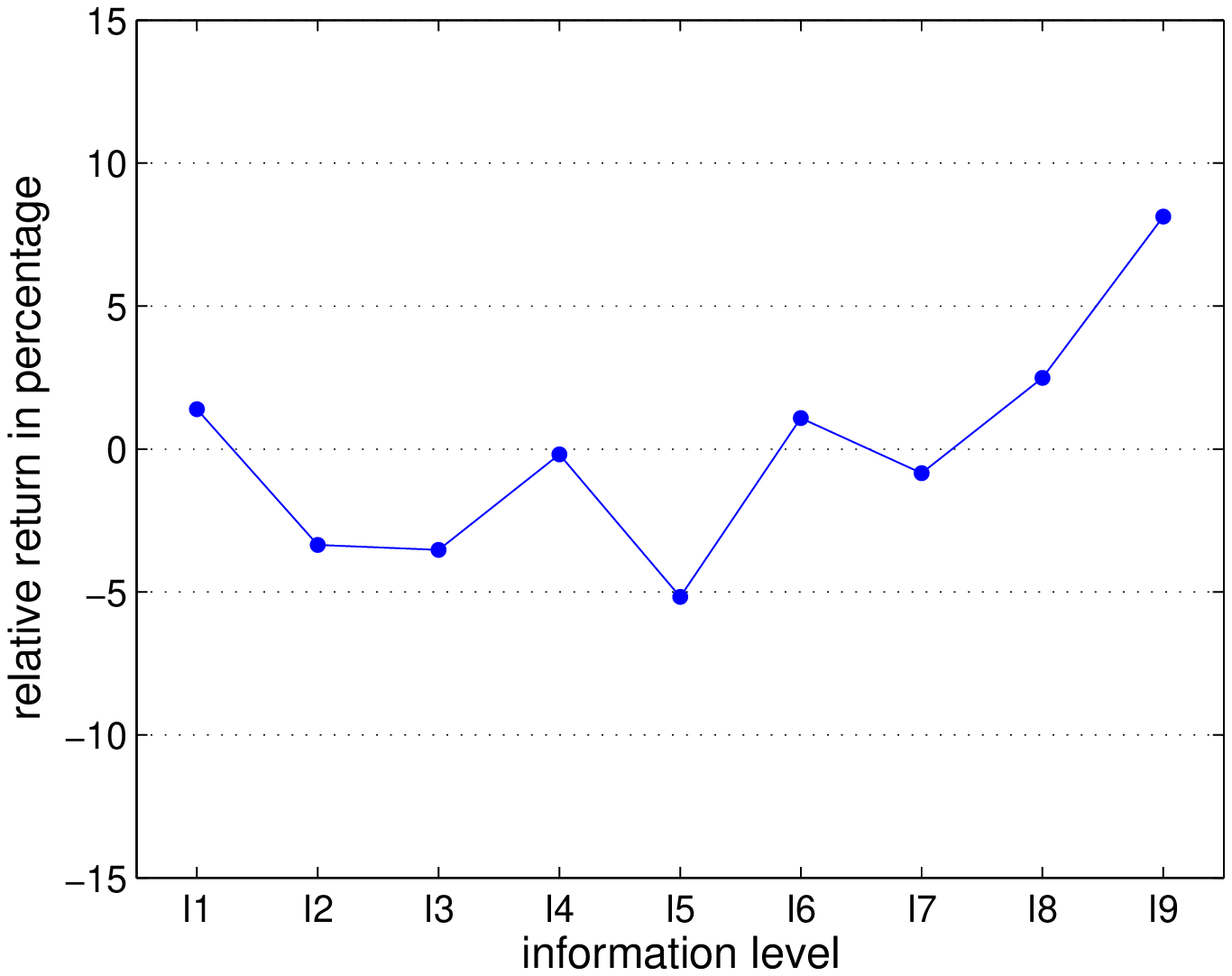}
}
\caption{Results of experiments (average of seven experiments). Return of traders relative to the market in percentage, as a function of the information. The returns are a non--monotonic function of information.}
\label{plot_experiment}
\end{center}
\end{figure}

The main interest is in how information affects the performance of traders.
The net return of traders compared to the market return as a function of the information level can be seen in Fig. \ref{plot_experiment}, the results are the average of the seven experiments. One can verify that the returns do not grow monotonically with increasing information. Traders having the first five levels of information do not outperform the average and only the best informed traders (insiders) are able to gain excess returns compared to the market \cite{huber.2006,huber.kirchler.sutter.2006}. For a statistical comparison of performance of traders we ran the Wilcoxon rank sum test for equal medians \cite{wilcoxon1,wilcoxon2}, on the relative performance for pairs of information levels. The p--values of the tests can be found in Table \ref{tab:exp_test}. Though in many of the cases the result of the test does not exclude the hypothesis of the returns being drawn from the same population, one can see that only the very well informed traders (\textit{I8} and \textit{I9}) perform significantly better than \textit{I3} and \textit{I5} on the \(0.05\) significance level, and the averagely informed (\textit{I5}) underperform the least informed (\textit{I1}) at the \(0.1\) significance level.

\begin{table}
\caption{p--values of the Wilcoxon rank sum test for equal medians on differences in performance between the information levels.
* significant at the \(0.05\) level
** significant at the \(0.1\) level}
\label{tab:exp_test}       
\begin{center}
\resizebox{0.95\columnwidth}{!}{
\begin{tabular}{lllllllllll}
    & I1 & I2 & I3  & I4 & I5 & I6 & I7 & I8 \\
\hline
I2 \vline & 0.710 &  &  & & & & &\\
I3 \vline & 0.210 & 0.460 &  &  & & & &\\
I4 \vline & 0.800 & 0.900 & 0.130 &  &  & & &\\
I5 \vline & 0.070** & 0.530 & 0.530 & 0.160 &  &  & &\\
I6 \vline & 1.000 & 0.900 & 0.070** & 1.000 & 0.210 &  &  &\\
I7 \vline & 1.000 & 1.000 & 0.530 & 0.620 & 0.320 & 0.800 &  &\\
I8 \vline & 0.530 & 0.620 & 0.040* & 0.260 & 0.020* & 0.900 & 1.000 &  &\\
I9 \vline & 0.210 & 0.260 & 0.010* & 0.130 & 0.010* & 0.320 & 0.320 & 0.320 &\\
\noalign{\smallskip}\hline
\end{tabular}
}
\end{center}
\end{table}

Since all information in the experiment is provided for free and is always correct, the result can not be due to information cost or wrong information. Furthermore, implementing an information cost in the system would possibly enlarge the disadvantage of being averagely informed: it would decrease returns for average and high information levels most.

A tool for corroborating the relevance of results in artificial markets to the real-world is analysing from the point of view of some of the well known empirical stylized facts of markets \cite{cont.2001}. While not getting stylized facts in a simulation can falsify the assumptions made, of course these facts in themselves do not confirm other results of the simulation.
The probability density function of price changes, the decay of the autocorrelation function of price changes and the decay of the autocorrelation function of absolute price changes were analysed in the experimental results. For the three tests the results showed similar results as data from real markets: the distribution of returns was fat tailed, the autocorrelation of returns decayed fast and the autocorrelation of absolute returns decayed slowly (volatility clustering) \cite{kirchler.huber.2006}.

\section{The simulations}\label{Section3}
We carried out computer simulations to numerically reproduce the results of the experiments done with human beings. The simulations were run using MATLAB\(^{\textregistered}\) programming language. 

\subsection{The market mechanism}\label{sim_mech}
In our simulation we programmed an essential double auction trading mechanism as it appears on most of real world financial markets, with a book containing the bid and ask orders. Since, in contrary to real world experiments, in a numerical simulation one has the possibility to analyse truly random traders, we implemented ten agents with different levels of information going from zero--information (random traders), \textit{I0} to \textit{I9}, and with the possibility of using different trading strategies as will be discussed in details in Section \ref{sim_info} and Section \ref{sim_strat}. 
In order to be able to estimate the error of our results we carried out \(10000\) runs in each simulation (100 sessions of 100 runs).\\
The simulation setup was very similar to the one in the experiments: the market contained a risky asset
(stock) and a risk free bond (cash). 
Before beginning the simulation an information level was assigned to each of the ten agents (nine informed and one uninformed), thus having one agent for every level.
Initially all agents were endowed with 1600 units of cash and 40 shares of stock with initial value of 40 units each. Trading consisted of \(30\) periods each lasting \(100\) simulation steps. At the beginning of each period new information was delivered to the agents according to their information level as we will discuss in Section \ref{sim_info}. 
At the end of each period a risk free interest rate was paid on the cash held by the agents, dividends were paid on the shares held by the agents (the risk free interest rate was \(r_{f}=0.01\), the risk adjusted interest rate \(r_{e}=0.05\)) and the book was cleared. We also carried out simulations without clearing the book and found that the clearing process does not make much difference in the results.\\
The dividend process (being the source of future information) was determined before the beginning of the trading.
Similarly to the experiments, the dividend process was a random walk of Gaussian steps:
\begin{eqnarray}
D(i)=D(i-1)+0.1N(0,1).
\label{eq_sim_div}
\end{eqnarray}
with \(D(0)=0.2\), where \(N(0,1)\) is a normal distribution with zero mean and unit variance (in case of $D(i)<0$, we took $|D(i)|$).
We are carrying out finite time simulations, so that short trends in the random walk can have important effects on the dividend process and by that on the information structure and the price formation on the market. When studying the performance of heterogeneously informed agents we carried out measurements with different dividend processes. The results shown are the statistics of 100 simulation sessions, each being run with a different random dividend process. One session consists of 100 simulation runs carried out with the same dividend process.

\subsection{Information}\label{sim_info}
Overall we implemented ten levels of information, a completely uninformed trader  (random trader), \textit{I0} and nine informed traders with different levels of information from \textit{I1} to \textit{I9}, where agent \textit{Ij} has information of the dividends for the end of the current period and of \((j-1)\) forthcoming periods (forecasting ability). The information received by traders was the present value of the stock conditional on the basis of their forecasting ability. This was determined by Gordon's formula (Eq. \ref{eq_gordon}). 

\subsection{Trading strategies}\label{sim_strat}
At the beginning of each period agents submit orders according to their idea of the value of stocks. After that, during the period, in every second, one trader is chosen randomly who either accepts a limit order from the book (gives a market order) or puts a new limit order to the book.

Since we do not have exact information on how traders use their information in real world and in the experiments, we gave the possibility to simulated traders to strictly apply the fundamental information they get (\textit{fundamentalists}), not to take any information in account except the current price, i.e. trade randomly (\textit{random traders}) or to look at other pieces of information such as trends (\textit{chartists}). In this paper we show results for the case of fundamentalist and random traders, these strategies are described below. The details of the trading strategies and order placing mechanisms can be found on the web page: \texttt{http://www.phy.bme.hu/\~{}bence}

\subsubsection{Fundamentalists}\label{sim_fund}
Fundamentalist traders strictly believe in the information they receive. If they find an \textit{ask order} with a price lower or a \textit{bid order} with a price higher than their estimated present value, i.e. \(E(V|I_{j,k})\), they accept the limit order, otherwise they put a new limit order between the former best bid and best ask prices.

\subsubsection{Random traders}\label{sim_rand}
Random traders put orders randomly. With probability \(0.5\) they put an ask (bid) order slightly higher (lower) than the current price.

\section{Results}\label{Section4}
In our simulations we focused on the effect of information on the performance of agents throughout the market session. We also analysed the results from the point of view of stylized facts of stock markets. In order to reduce statistical errors we carried out 10000 runs of the simulation.

\subsection{Final wealth as a function of information}\label{wealth_info}
\begin{figure}
\begin{center}
\resizebox{0.75\columnwidth}{!}{%
  \includegraphics{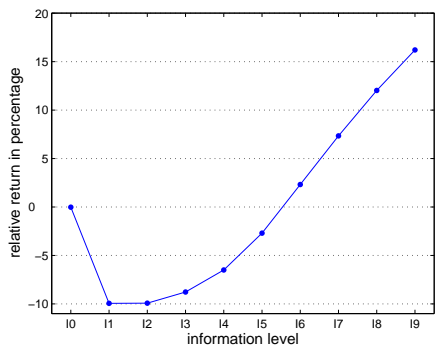}
}
\caption{Results of simulations (average of 10000 runs). Returns of traders relative to the market in percentage, as a function of information. One can see that having average level of information is not necessarily an advantage.}
\label{plot_wealth}
\end{center}
\end{figure}

The final return relative to that of the whole market can be seen in Fig. \ref{plot_wealth}, the results are the average of \(100\) sessions, each session consisting of 100 runs. The results are in good accordance with the experimental results: we get a curve we call \textit{J--curve}. The agents having average level of information (\textit{I1--I5}) perform worse than the completely uninformed random agent (\textit{I0}). The best informed agents outperform the market. (Besides common sense, the latter can be justified also using mathematics, since compared to the market index the simulation is a zero sum game. If the non--informed gets more or less the market return and the averagely informed are losers, then the well--informed must get excess gain.). To test the hypothesis of the J--curve we ran the Wilcoxon rank sum test for equal medians \cite{wilcoxon1,wilcoxon2}, on the relative performance for pairs of information levels. The p--values of the tests can be found in Table \ref{tab:sim_test}. One can see that the hypothesis of returns for different information levels being drawn from the same population can be excluded in almost all cases at the \(0.05\) significance level.
This result and its relevance to real markets will be discussed further in Section \ref{conclusions}.

\begin{table}
\caption{p--values of the Wilcoxon rank sum test for equal medians on differences in performance between the information levels.
* significant at the \(0.05\) level
** significant at the \(0.1\) level}
\label{tab:sim_test}       
\begin{center}
\resizebox{0.95\columnwidth}{!}{
\begin{tabular}{lllllllllll}
    & I0 & I1 & I2 & I3  & I4 & I5 & I6 & I7 & I8 \\
\hline
I1 \vline & 0.000* &  & & & & & & & & \\
I2 \vline & 0.000* & 0.932  &  & & & & & & & \\
I3 \vline & 0.000* & 0.507  & 0.385  &  & & & & & & \\
I4 \vline & 0.000* & 0.003* & 0.002* & 0.013* &  & & & & & \\
I5 \vline & 0.000* & 0.000* & 0.000* & 0.000* & 0.000* &  & & & & \\
I6 \vline & 0.144  & 0.000* & 0.000* & 0.000* & 0.000* & 0.000* &  & & & \\
I7 \vline & 0.000* & 0.000* & 0.000* & 0.000* & 0.000* & 0.000* & 0.000* &  & & \\
I8 \vline & 0.000* & 0.000* & 0.000* & 0.000* & 0.000* & 0.000* & 0.000* & 0.009* &  & \\
I9 \vline & 0.000* & 0.000* & 0.000* & 0.000* & 0.000* & 0.000* & 0.000* & 0.001*  & 0.057**  &  \\
\noalign{\smallskip}\hline
\end{tabular}
}
\end{center}
\end{table}

\begin{figure}
\begin{center}
\resizebox{0.75\columnwidth}{!}{%
  \includegraphics{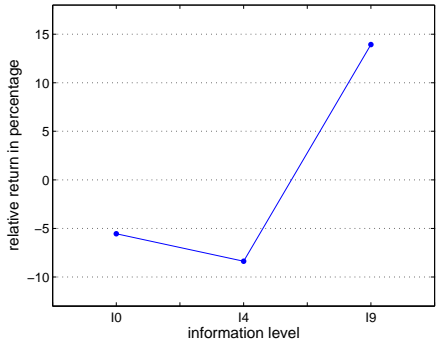}
}
\caption{Results of simulations (average of 100 runs). Returns of traders relative to the market in percentage, as a function of information. Already, in case of 3 agents one can identify the J--curve.}
\label{3traders}
\end{center}
\end{figure}

To understand why the random trader gets almost exactly the market return and to see how the relative wealth of agents looks like for simpler cases, we ran simulations with only three agents in the market (in this case with the standard deviation of dividends being 0.01, $r_f=0.001$ and $r_e=0.005$): an uninformed (\textit{I0}), an averagely informed (\textit{I4}) and a well informed (\textit{I9}).
Fig. \ref{3traders} shows the plot for this case similarly to Fig. \ref{plot_wealth}. We can exclude the monotonicity of the curve and even if with three points it is harder to call it a J--curve, we can see that the random trader performs better than the average informed one and only the well--informed gets excess returns. In this case also the random trader performs under the market level, giving an explanation for the question raised: in case of enough actors present on the market, the price impact of the random trader becomes negligible, thus the random trader has equal probability of being beaten by the market and of beating the market. To have more insight into this process, in Table \ref{tab:random_perf} we show the relative performance of the random trader in case of different number of agents (when always the least informed agents are present on market).

\begin{table}
\caption{Relative performance of the random trader on markets with different number of agents. Note that in this case always the least informed are present on the market (e.g. in case of 3 traders: \textit{I0}, \textit{I1} and \textit{I2}). As more agents are present on the market, the performance of the random agent approaches market return.}
\label{tab:random_perf}       
\begin{center}
\resizebox{0.95\columnwidth}{!}{
\begin{tabular}{llllll}
number of traders     &  3 &  5 &  7 &  9  &  10 \\
\hline
relative performance  & & & & & \\of random trader [\%]  & -2.7 & -1.0 & -0.2 & +0.2 & -0.1 \\
\noalign{\smallskip}\hline
\end{tabular}
}
\end{center}
\end{table}

\subsection{Stylized facts}\label{stylized}

\begin{figure}
\begin{center}
\resizebox{0.75\columnwidth}{!}{%
  \includegraphics{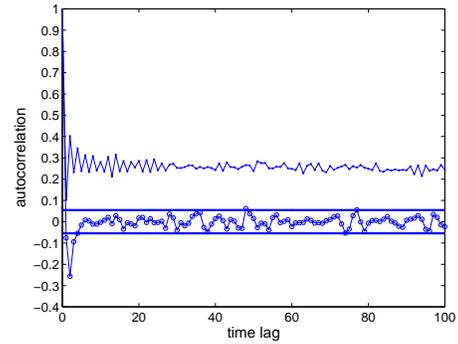}
}
\caption{Autocorrelation functions of returns (circles and lines) and absolute returns (dots and lines) and the noise level (solid lines). Autocorrelation of returns decays fast under the noise level while autocorrelation of absolute returns decays very slowly, showing the clustering of volatility. Results of one separate run of the simulations.}
\label{plot_autocorr}
\end{center}
\end{figure}

We analysed the results of our simulations from the point of view of the three common empirical stylized facts as were done for the experiments \cite{kirchler.huber.2006}. Fig \ref{plot_autocorr} shows the autocorrelation functions of returns (circles and lines) and of absolute returns (dots and lines). The noise level of the computations is also included in the plot (straight lines). One can see that the autocorrelation of returns decays fast under the noise level (with a negative overshoot for small lags as it is usual in real world markets too), thus there is no long time correlation in price changes. On the other hand the autocorrelation of absolute returns decays slowly showing the fact that big price changes tend to cluster (volatility clustering). (A slight even--odd oscillation is visible in the autocorrelation of absolute returns, this is an artifact of our simulation process, as there are many cases in which the intertrade time is two simulation steps, resulting in this oscillation.) 

\begin{figure}
\begin{center}
\resizebox{0.75\columnwidth}{!}{%
  \includegraphics{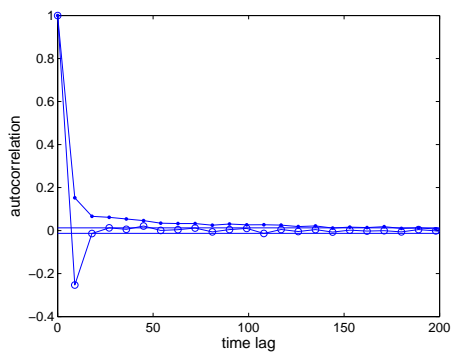}
}
\caption{Autocorrelation functions of returns (circles and lines) and absolute returns (dots and lines) and the noise level (solid lines) for the logarithmic returns of General Electric. Autocorrelation of returns decays much faster than that of the absolute returns, in agreement with the simulation results.}
\label{GE}
\end{center}
\end{figure}

For comparison, in Table \ref{tab:moments} we have reported the first four moments of the log-return distribution for tick-by-tick General Electric (GE) prices in October 1999 (55559 data points). In Figure \ref{GE}, for these data, the autocorrelation of signed and absolute log-returns is plotted as in Figure \ref{plot_autocorr}. From Table \ref{tab:moments} it can be seen that the distribution function of absolute returns in the simulation is leptokurtic, similarly to real world data, even if the kurtosis is
lower. Running the Jarque--Bera test, for goodness-of-fit to a normal distribution \cite{jarquebera}, we can rule out the normality of the distribution of the absolute returns for both cases.

\begin{table}
\caption{The first four moments of the logarithmic return distribution for tick-by-tick General Electric (GE) prices and for the simulated prices.}
\label{tab:moments}       
\begin{center}
\resizebox{0.95\columnwidth}{!}{
\begin{tabular}{llll}
   &  GE data &  simulation\\
\hline
mean  & 2.1215e-06  & 3.5221e-05 & \\
standard deviation & 4.0078e-04  & 0.0250 & \\
skewness & -0.0698  & 0.1233 & \\
kurtosis & 36.2677  & 7.9541  & \\
\noalign{\smallskip}\hline
\end{tabular}
}
\end{center}
\end{table}

When testing for the stylized facts, we also studied markets with only random agents trading and we found similar stylized facts; thus we can state that these empirical facts are effects mainly due to the continuous double auction trading mechanism as it has been mentioned before in Ref. \cite{LiCalzi.Pellizzari.2002}.

\section{Conclusions} \label{conclusions}
In this paper we presented a model of an experimental and a simulated double auction stock market with cumulative information delivered to traders. We focused on the value of information for the traders, in case of several information levels. 

The results of the experiments and the simulation show a non trivial, non monotonic dependence of agents' returns on the amount of information possessed. We found, that averagely informed traders perform worse than the market level. In the simulations we analysed the case of non informed traders and found that if there are enough traders present on the market, the non informed, random trader is able to get the market return. Hence we can state that averagely informed traders perform worse than the completely non informed, thus in case of the averagely informed traders the information has a negative effect on the performance.
Only the most informed traders (insiders) are able to gain above--average returns.

These results can give a possible explanation for a puzzling real life phenomenon.  Most of the professional fund managers on stock markets perform worse on the long run than the market itself, i.e. they get lower returns than a random trader would get in the same period, see e.g. Ref. \cite{malkiel.2003b}. The possible cause for this bad performance can be seen from our results: most of the professional fund managers are not insiders neither completely uninformed. They fit into the middle of our curve on Fig. \ref{plot_wealth}. Traders taking random decisions can outperform them on the long run, receiving the market return.
The reason for this phenomenon can be interpreted in the following way: traders having no forecasting ability trade randomly and can not be exploited by other traders. At the same time, traders having average forecasting horizon but believing in the information they possess, can be exploited by better informed traders, insiders.
Of course the behaviour of real world traders is much more complicated than the ones implemented in our simulations, e.g. they have the possibility of modifying strategy, switching between stocks or sectors whereas in our experimental and simulation platform only one stock was present. Nevertheless the non-monotonic behaviour of Fig. \ref{plot_experiment} and Fig. \ref{plot_wealth} suggests an explanation for the low average performance of actively managed funds.

It is important to stress, that while heterogeneous beliefs of agents are necessary for trading (if all agents had the same expectations, no one would find it attractive to trade), we were able to reproduce the J--curve of the experiments in our simulations by implementing fundamentalist strategy. Thus it is enough to assume that
traders use the information they possess to get the non monotonic relationship of net returns of traders as a function of information levels.

\section*{Acknowledgments}
We would like to thank J\'anos Kert\'esz, Sorin Solomon and Gilles Daniel for discussions and for raising some interesting questions. Support by OTKA T049238 is acknowledged.
E. S. was partially supported by the Italian MIUR grant "Dinamica di altissima frequenza nei mercati finanziari".

%

\end{document}